\documentclass[a4paper,twocolumn,superscriptaddress,11pt,accepted=2020-06-04]{quantumarticle}
\pdfoutput=1
\usepackage[numbers,sort&compress]{natbib}
\usepackage{epsfig}
\usepackage{amsfonts}
\usepackage{amsmath}
\usepackage{amssymb}
\usepackage{amsthm}
\usepackage{color}
\usepackage{multirow}
\usepackage[normalem]{ulem}
\newcommand{\stkout}[1]{\ifmmode\text{\sout{\ensuremath{#1}}}\else\sout{#1}\fi}
\usepackage{latexsym}
\usepackage{mathrsfs}
\usepackage{natbib}
\usepackage{verbatim}
\usepackage[T1]{fontenc}
\usepackage{float}
\usepackage{rotating}

\usepackage[colorlinks=true,linkcolor=blue,citecolor=magenta,urlcolor=blue]{hyperref}

\DeclareMathOperator{\Tr}{tr}

\newcommand{\ket}[1]{|#1\rangle}

\newcommand{\ketbra}[2]{|#1\rangle\langle#2|}

\begin{document}


\title{Does violation of a Bell inequality always imply quantum advantage in a communication complexity problem? }


\author{Armin Tavakoli}
\affiliation{D\'epartement de Physique Appliqu\'ee, Universit\'e de Gen\`eve, CH-1211 Gen\`eve, Switzerland}

\author{Marek \.Zukowski}
\affiliation{International Centre for Theory of Quantum Technologies (ICTQT), University of Gdansk, 80-308 Gdansk, Poland}

\author{\v{C}aslav Brukner}
\affiliation{Faculty of Physics, University of Vienna, Boltzmanngasse 5, A-1090 Vienna, Austria}
\affiliation{Institute of Quantum Optics and Quantum Information, Austrian Academy of Sciences, Boltzmanngasse 3, A-1090 Vienna, Austria}

\begin{abstract}
Quantum correlations which violate a Bell inequality are presumed to power better-than-classical protocols for solving communication complexity problems (CCPs). How general is this statement? We show that violations of correlation-type Bell inequalities allow advantages in CCPs, when communication protocols are tailored to emulate the Bell no-signaling constraint (by not communicating measurement settings). Abandonment of this restriction on classical models allows us to disprove the main result of, inter alia, [\href{https://doi.org/10.1103/PhysRevLett.89.197901}{Brukner et. al., Phys Rev. Lett. 89, 197901 (2002)}]; we show that quantum correlations obtained from these communication strategies assisted by a small quantum violation of the CGLMP  Bell inequalities do not imply advantages in any CCP in the input/output scenario considered in the reference. More generally, we  show that there exists quantum correlations, with nontrivial local marginal probabilities, which violate the $I_{3322}$ Bell inequality, but do not enable a quantum advantange in any CCP, regardless of the communication strategy employed in the quantum protocol, for a scenario with a fixed number of inputs and outputs
\end{abstract}


\maketitle

\section{Introduction} Entanglement in itself cannot be used for information transfer.  However, when combined with classical communication, it becomes a paradigmatic resource for quantum information transfer. It can amplify the capacity of a channel \cite{Capacities, Capacities2}, most famously in superdense coding \cite{DenseCoding}. Also, it can be used as a resource for better-than-classical communication complexity \cite{CB97, Brassard}. Such reductions of communication complexity have a range of applications (see e.g.~\cite{Gavinsky, RSP, FakeTriangle, Laplante, Bridge}) and are central tools for understanding the power of entanglement as a resource, both in terms of the extent to which it can outperform classical approaches \cite{Buhrman98, Raz, Tapp} and how it compares to other quantum resources \cite{PZ10, TP17, Sergii}.

Communication complexity problems (CCPs) are tasks in which separated parties collaborate to compute a function dependent on inputs distributed among them, while only being allowed a limited amount of communication\footnote{An alternative approach to CCP considers the minimal amount of communication required to compute a function with distributed inputs. However, in this work, our focus is scenarios in which tasks are performed with limited communication.}. In their simplest form, such tasks can be viewed as games in which two parties Alice and Bob hold random inputs $X$ and $Y$ respectively and collaborate so that one of them (say Bob) can compute a function $f(X,Y)$. Alice communicates a classical message $m(X)$ to Bob who outputs a guess $g(m,Y)$ for the value of $f$. If the guess is correct, the partnership earns a ``point". Importantly, the communication is limited so that the alphabet of $m$ is smaller than that of $X$, typically rendering perfect evaluations  of $f$  impossible. The CCP is to find for Alice and Bob a communication strategy maximising the score, i.e.~the averaged (over the distribution of inputs)  number of points.

By sharing entanglement, Alice and Bob can sometimes increase their score beyond what is classically achievable \cite{CB97}. To this end, it is necessary that they exploit entanglement to distribute strong correlations that violate a Bell inequality. We illustrate this with an example \cite{BC01}. Alice (Bob) has fully random inputs $X=(x_0,x)\in[2]^2$ ($Y=y\in[2]$), where $[s]$ denotes the set $\{0,\ldots,\lvert s\rvert-1\}$.  The CCP (game) is: Bob  earns a point if he gives his guess $g$ equal to $f=x_0+xy\mod{2}$, while Alice can send him only a binary (bit) message $m(x_0,x)$.  The score in the CCP is written  $\mathcal{S}=1/8\sum_{x_0,x,y} P(g=f|x_0,x,y)$.  The optimal classical score, which is $\mathcal{S}=\frac{3}{4}$, is achieved with some deterministic encoding/decoding procedure. Due to the small number of inputs, one can easily consider all possible messaging and guessing strategies. One finds that there are several different strategies achieving the optimal classical score in the CCP. It can be shown, see \cite{CB97} and e.g. \cite{SINGLE}, that among other ones, there is an optimal strategy which runs as follows: Alice sends $m(x_0,x)=a(x)+x_0\mod{2}$ and Bob guesses $g=m+b(y) \mod{2}$, where $a(x)$ and $b(y)$ are binary-valued functions of the inputs $x$ and $y$ respectively. The winning condition $f=g$ now reduces to $a(x)+b(y)=xy$, which allows us to put the score in the form of the Clauser-Horne-Shimony-Holt (CHSH) \cite{CHSH} Bell inequality: $\mathcal{S}=1/4\sum_{x,y} P(a+b=xy|x,y)\leq 3/4$. Thus immediately  the following entanglement-assisted strategy becomes relevant: Alice and Bob use their inputs $(x,y)$ as settings in a quantum test of the CHSH inequality where $a,b\in[2]$ are their respective local outcomes. Having obtained her outcome in the CHSH test, Alice sends the message  $m(x_0,x)=a+x_0\mod{2}$ to Bob. Notice that this message emulates the Bell no-signaling constraint in the sense that it does not allow Bob to read out the value of $x$, which was used as a setting in the CHSH test. Bob uses his outcome in the CHSH test to construct the guess  $g=m+b \mod{2}$.  Since shared entanglement enables Alice and Bob to violate the CHSH inequality, this quantum strategy leads to a score of up to $\mathcal{S}=\frac{1}{2}+ \frac{1}{2\sqrt{2}}\approx 0.854$, which is the largest possible violation of the CHSH inequality. Thus, the entanglement-assisted strategy holds an advantage over all possible classical strategies in the CCP.

There are many more results showing that every probability distribution that violates  \textit{specific} Bell inequalities has the ability of enhancing a CCP beyond  classical protocols. Examples include the Mermin  inequalities \cite{Mermin, CB97}, the Collins-Gisin-Linden-Massar-Popescu (CGLMP)  inequalities \cite{CGLMP, BZ02, BP03, TP17}, the elegant Bell inequality \cite{elegant}, Bell inequalities for random access coding \cite{PZ10, WO10, TM16, HS17}, the biased CHSH inequalities \cite{Sandu, Bridge} and a large class of bipartite many-outcome Bell inequalities \cite{TZ17}. More generally, Ref.~\cite{BZ04} showed that the violation of \textit{every} multipartite correlation Bell inequality with binary outcomes implies beating the best possible classical score  in a corresponding CCP constructed  by generalising the above example based on the CHSH inequality. See the topical review~\cite{BC10} for further discussions. This fauna of results begs the question: does every nonlocal probability distribution  (i.e.~a probability distribution that violates a Bell inequality) lead to an advantage in a CCP? To show such advantages, one requires only an example of a CCP in which access to the nonlocal probability distribution is advantageous. However,  proving that no such advantage is possible is significantly more challenging; one must rule out the possibility of an advantage in \textit{every} possible CCP, i.e.~no matter the number of inputs and outputs, the choice of score and the chosen classical communication strategy.

Whereas we do not provide a decisive answer to whether Bell nonlocality always implies advantages in CCPs, we show that there exists a natural input/output scenario in which Bell nonlocality does not enable a quantum advantage in any CCP. We first formalise classical and entanglement-assisted CCPs. Then, we show how to map multipartite $d$-outcome correlation Bell inequalities to corresponding CCPs. This method allows to e.g. reproduce the examples studied in the literature, listed in the previous paragraph. We prove that a violation   of a Bell inequality, together with restricted communication strategies, which do not reveal the input the sender would use to define her measurement setting the Bell inequality test, implies beating an analogous classical protocol for the corresponding CCP.

 This restriction on classical strategies is tacitly used in several previous works (see e.g.~Refs.~\cite{BZ02, BP03, TP17}) which enables a quantum advantage. Our  more complete analysis of classical strategies no longer sustains the generality of e.g.~the main result of Ref.~\cite{BZ02}, that every violation of the CGLMP inequality combined with the above mentioned communication strategies implies an advantage in some CCP for a fixed number of inputs and outputs. This leads us to consider the classical simulation of entanglement-assisted CCPs. We consider a situation with fixed number of inputs and outputs and show that there exists a quantum nonlocal probability distribution that does not enable better-than-classical communication complexity, regardless of the communication strategy and the choice of score. Our results are in opposition to the common belief that Bell nonlocality always is useful for better-than-classical communication complexity.

\section{Formal scheme of the communication complexity problems analysis}
We mainly consider two-party protocols. These are formulated as {\em games}. Alice and Bob each receive random inputs, respectively  $X\in[N_\text{A} ]$ and $Y\in[N_\text{B}]$. Alice sends a message $m\in [ M]$ (with $M<N_\text{A}$) to Bob who outputs $g\in [G]$, which is awarded with $t^g_{X,Y}$ points. The tuple $(N_\text{A},N_\text{B},M,G)$ corresponds to a choice of \textit{scenario}. The score of a specific CCP within the chosen scenario is written as
\begin{equation}\label{effi}
	\mathcal{S}[p(g|X,Y)]=\sum_{g,X,Y} t_{X,Y}^g p(g|X,Y)
\end{equation}
where  $p(g|X,Y)$ is the probability of Bob's output for local inputs $X,Y$. Notice that the scoring function can always absorb prior probabilities $p(X,Y)$.

In a classical picture, Alice encodes her message with a function $E:[N_\text{A}]\rightarrow [M]$ and Bob constructs his guess with a function $D:[M]\times [N_\text{B}]\rightarrow [G]$. The choice of $(E,D)$ can be coordinated via a shared random variable $\lambda$, with some probability distribution $p(\lambda)$. Therefore, a classical model is of the form
\begin{equation}\label{classical}
p^\text{C}(g\lvert X,Y)=\sum_{\lambda} p(\lambda)p_\lambda(g|X,Y).
\end{equation}
where the deterministic distribution is $p_\lambda(g|X,Y)=\sum_m p(m\lvert X,\lambda)p(g\lvert m,Y,\lambda)$ with  $p(m\lvert X,\lambda)=\delta_{m,E_\lambda(X)}$ and $p(g\lvert m,Y,\lambda)=\delta_{g,D_\lambda(m,Y)}$. Due to linearity in Eq.~\eqref{effi}, the largest score is found with a deterministic communication strategy. We therefore define the optimal classical  score in a CCP as 
\begin{equation}\label{cbound}
\mathcal{S}^\text{C}=\max_{\lambda} \mathcal{S}[p_\lambda(g|X,Y)].
\end{equation}

In contrast, if Alice and Bob share an entangled state $\rho$, they may use their inputs to select measurement settings  with associated outcomes $a$ and $b$ respectively. The statistics reads $p(a,b|X,Y)=\Tr\left[A_X^a\otimes B_Y^b \rho\right]$ where $A_X^a$ and $B_Y^b$ are measurement operators. Subsequently, Alice sends $m=E(a,X)$ for some function $E:[|a|]\times [N_\text{A}]\rightarrow [M]$ and Bob guesses $g=D(m,b,y)$ for some function  $D:[M]\times [|b|]\times [N_\text{B} ]\rightarrow [G]$,  where $|a|$ and $|b|$ denote the cardinality of the respective output spaces. Here, we have assumed that the Bell inequality test is performed before Bob receives Alice's message (in line with space-like separation). Moreover, although shared randomness could be absorbed into the shared entangled state, we treat it separately in order to emphasise that it is a classical resource. Therefore, a quantum model is of the form
\begin{equation}\label{quantum}
p^{\text{Q}}(g\lvert X,Y)=\sum_{\lambda} p(\lambda) p^\text{Q}_\lambda (g|X,Y),
\end{equation}
where 
\begin{multline}\nonumber
p^\text{Q}_\lambda (g|X,Y)=\\ \sum_{a,b,m} p(a,b\lvert X,Y)p(m\lvert a,X,\lambda)p(g\lvert m,b,Y,\lambda).
\end{multline}

\section{All violations of correlation Bell inequalities power advantages in constrained CCPs} Consider a Bell scenario with $N$ parties $O_1,\ldots, O_N$ who perform measurements labelled $x_1,\ldots, x_N$ with outcomes $o_1,\ldots,o_N\in[d]$ for some $d\geq 2$, and any Bell inequality of the  form  
\begin{equation}\label{Bell}
\mathcal{B}=\sum_{\vec{x}}\sum_r t_{\vec{x}}^{r}P_{\vec{x}}\left(\sum_{i=1}^N o_i=f_{\vec{x}}^r\right)\stackrel{\text{LHV}}{\leq} C,
\end{equation}
where $\vec{x}=(x_1,\ldots,x_N)$, $C$ is the LHV bound, $r$ is an additional integer-valued indexing parameter which allows for more general Bell inequalities\footnote{For instance the CHSH inequality requires only one value of $r$ but in order to write the CGLMP inequalities on the form \eqref{Bell} one requires several values of $r$.}, $f_{\vec{x}}^r\in[d]$ and $t_{\vec{x}}^{r}$ are real coefficients. The relation $\sum_i o_i=f_{\vec{x}}^r$ is evaluated modulo $d$. The Bell inequalities in Eq~\eqref{Bell} are sometimes referred to as full correlation Bell inequalities. To map \eqref{Bell} to a CCP, let $O_i$ (for $i=1,\ldots, N-1$) have an input $X_i=(x_i,x_0^{(i)})$ where $x_0^{(i)}\in[d]$. Each of these parties may send a message $m_i\in [d]$ after which $O_N$, who has input $X_N=x_N$, produces a guess $g\in[d]$ and earns  $t_{\vec{x}}^{r}/d^{N-1}$ points whenever $g=f_{\vec{x}}^r+\sum_{i=1}^{N-1} x_0^{(i)}$. The  score is defined as
\begin{equation}\label{CCP}
\mathcal{S}=\frac{1}{d^{N-1}}\sum_{\vec{x},\vec{x}_0} \sum_r t_{\vec{x}}^{r}P_{\vec{x}}\left(g=f_{\vec{x}}^r+\sum_{i=1}^{N-1} x_0^{(i)}\right),
\end{equation} 
where $\vec{x}_0=(x_0^{(1)},\ldots,x_0^{(N-1)})$. Notice that the coefficients $t_{\vec{x}}^{r}/d^{N-1}$ in the CCP do not depend on $\vec{x}_0$. To put the Bell inequality and the CCP on equal footing, let the $N$ parties share an entangled state and use their inputs $\vec{x}$ to perform a measurement with outcome $o_i\in[d]$. Then, the parties $O_i$ for $i=1,\ldots, N-1$ send $m_i=o_i+x_0^{(i)}$ to $O_N$ who outputs $g=o_N+\sum_i m_i$.  Here, in analogy with the previous CHSH-inspired example, the addition of $x_0^{(i)}$ in the message ensures that  $O_N$ cannot learn the input $x_i$ which was used as a setting in a Bell inequality test. Also, notice that the parties $O_1,\ldots,O_{N-1}$ only use part of their inputs for choosing a measurement setting. This is in analogy with previous litterature (e.g.~Refs.~\cite{BZ02, BZ04, PZ10, TZ17}). We then find $\mathcal{S}=\mathcal{B}$. In comparison, consider a classical situation that is restricted to the same type of communication strategies, i.e.~``additive'' messages on the form $m_i=o_i(x_i)+x_0^{(i)}$, where $o_i(x_i)$ is a function of $x_i$, that do not reveal the value of $x_i$. Naturally, this leads to $\mathcal{S}=\mathcal{B}\leq C$. Therefore, when restricting to additive communication strategies for both the quantum and classical situation, one finds that violation of Eq.~\eqref{Bell} implies $\mathcal{S}>C$, i.e.~a quantum advantage in the CCP. The above construction generalises results in Refs.~\cite{CB97, BZ02, BP03, TP17, PZ10, TM16, HS17, Bridge, BZ04}. Communication which does not allow one to reveal measurement settings (as above) is important in scenarios in which the task function should be calculated in a way which does not allow  an eavesdropper to learn the inputs of a sender, or even in a more subtle situation in which a sender does not want the receiver to know her inputs.

For $d=2$, the scenario reduces to that of Ref.~\cite{BZ04}, in which it was shown that messages of the form $m_i=o_i+x_0^{(i)}$ lead to the optimal classical score \eqref{cbound}. However, the same does not have to be true for $d>2$. We shall explicitly show that such is not the case using the specific example of Ref.~\cite{BZ02}. Ref.~\cite{BZ02} showed  (via the above map) that every violation of the CGLMP inequality \cite{CGLMP} implies an advantage in a corresponding CCP in which the communication is restricted to the  additive communication strategies defined above. As we show next, this constraint effectively excludes the optimal classical strategy.

\section{The CGLMP inequality and communication complexity}
Let us consider the CCP of Ref.~\cite{BZ02} obtained by choosing \eqref{Bell} as the CGLMP inequality. This inequality is a facet Bell inequality when Alice and Bob have two settings $x,y\in[2]$ and three possible outcomes $a,b\in[3]$;
\begin{multline}\label{cglmp}
\mathcal{B}_\text{cglmp}=\\
\frac{1}{4}\sum_{x,y} \big[P_{xy}(a+b=f^1)-P_{xy}(a+b=f^2)\big]\stackrel{\text{LHV}}{\leq} \frac{1}{2},
\end{multline}
where $f^1=-xy$ and $f^2=-xy+(-1)^{x+y}$. Using two entangled qutrits, one can reach the maximal quantum violation $\mathcal{B}^{\text{Q}}_\text{cglmp}\approx 0.7287$ \cite{Miguel}. In the corresponding CCP, Alice (Bob) has 
random inputs $x_0\in[3]$ and $x\in[2]$ ($y\in[2]$). Alice may send a ternary message $m\in[3]$ to Bob who outputs a guess $g\in[3]$. The score \eqref{CCP} is  
\begin{multline}\label{score}
\mathcal{S}_\text{cglmp}=\\
\frac{1}{12}\!\sum_{x_0,x,y}\!\left[ P_{xy}(g=x_0+f^1)-P_{xy}(g=x_0+f^2)\right].
\end{multline}
For an additive communication strategy, the violation of \eqref{cglmp} is necessary and sufficient for outperforming the corresponding classical value $\mathcal{S}_\text{cglmp}=1/2$. Ref.~\cite{BZ02} restricted itself to such additive communication strategies (for both classical and quantum models) to prove that a quantum enhancement of classical protocols in this CCP is possible if and only if the CGLMP inequality is violated. Since additive communication strategies assisted by Bell nonlocal correlations allow for a strong link between CCPs and Bell inequality violations, it appears natural that such communication strategies would power quantum-over-classical advantages in CCPs. However, as we show next, this does not necessarily mean that the best classical strategies are of the additive type, i.e.~it is not always optimal to tailor a classical strategy to a Bell inequality.

We now relax the assumption of additive communication strategies ($m=x_0+a \mod{3}$) and compute the optimal classical score \eqref{cbound}. There are $3^{12}$ deterministic encoding and guessing strategies. Interestingly, by separately considering all of them, we find that
\begin{equation}\label{ebound}
\mathcal{S}^\text{C}_\text{cglmp}=\frac{2}{3}.
\end{equation}
This can be saturated by Alice sending $ m(x_0,x)=\delta_{x,0}\delta_{x_0,2}+2\delta_{x,1}\delta_{x_0,1} \mod{3}$ and Bob guessing $g(m,y)=2\delta_{y,0}m+\delta_{y,1}\left(m+1\right) \mod{3}$. 
 Note that $m=1,2$ informs Bob of the value of $x$. 
 
Thus, the broad class of communication strategies considered in Ref.~\cite{BZ02} is insufficient to find the optimal classical score. A large violation of the CGLMP inequality,
$\mathcal{B}_\text{cglmp}>2/3$, indeed does imply an advantage over general classical protocols for the CCP. However, weaker violations, $1/2<\mathcal{B}_\text{cglmp}\leq 2/3$, are insufficient to achieve the same feat. In Appendix.~\ref{AppA} we show that analogous criticism applies to the CCPs of Refs.~\cite{BP03, TP17}. Also, we have numerically considered whether the limitation $\mathcal{S}_\text{cglmp}\leq \mathcal{B}_\text{cglmp}^\text{Q}$ can be overcome by using a more general message (see Appendix.~\ref{AppB} for details). However, we have found no improvement over the strategy in which Alice and Bob maximally violate the CGLMP inequality and the message is additive.

Departing from the particular CCP \eqref{score}, we remind ourselves that entanglement-assisted advantages originate from the probability distribution $p(g|x_0,x,y)$. Evidently $p(g|x_0,x,y)$ may lack a classical model even when the \textit{specific} score \eqref{score} does not exceed the classical bound \eqref{ebound}. Is it the case that for every probability distribution $p(g|x_0,x,y)$ obtained by a trit-communication and a violation of the CGLMP inequalities, there exists some other CCP in which a higher score is obtained than is classically possible? Presently, we answer this for the case of the additive communication strategy. Let Alice and Bob use their shared entanglement to generate a probability distribution of the form   
\begin{equation}\label{dist}
p(a,b|x,y)=vp^\text{cglmp}(a,b|x,y)+\frac{1-v}{9},
\end{equation} 
where $p^\text{cglmp}(a,b|x,y)$ maximally violates the CGLMP inequality and $v\in[0,1]$ is the protocol visibility parameter. This violates the CGLMP inequality when $v>0.6861$. The probability distribution $p_v^\text{Q}(g|x_0,x,y)$, obtained via shared entanglement and an additive communication strategy,  beats the classical bound \eqref{ebound} when $v>0.9149$. We seek the largest $v$ for which $p_v^\text{Q}$ can be simulated by a classical model. This means solving the linear program
\begin{equation}\label{LP}
\begin{array}{rcl}
\max\limits_{p(\lambda)} v && \text{s.t.} \qquad  p(\lambda)\geq 0, \qquad \sum_{\lambda}p(\lambda)=1,\\[7pt]
\text{and} && p_v^\text{Q}(g|x_0,x,y)=\sum_{\lambda} p(\lambda)p_\lambda(g|x_0,x,y).
\end{array}
\end{equation}
By considering $p_\lambda(g|x_0,x,y)$ for all possible deterministic strategies, we have found that the corresponding polytope of classical probability distributions has  $47601$ vertices. We have evaluated the linear program and found $v\approx0.7943$. Hence, probability distributions $p_v^\text{Q}(g|x_0,x,y)$ for $0.7943<v\leq 0.9149$ indeed imply an advantage over classical protocols in some CCP despite the particular CCP \eqref{score} failing to detect it. However, when $0.6861<v\leq 0.7942$ the CGLMP inequality is violated, but the probability distribution $p_v^\text{Q}(g|x_0,x,y)$ can be classically modelled.

\section{Bell nonlocality without CCP advantages in fixed scenarios} The above classical simulation focuses on entanglement-assisted correlations obtained via an additive communication strategy.  Here, we prove a more general statement: that for a given scenario (that is, a fixed number of inputs and outputs) there exists a nonlocal  probability distribution which  cannot be used to improve any CCP beyond classical constraints, regardless of the choice of communication strategy. Specifically, we find a nonlocal probability distribution that when combined with \textit{any} communication strategy gives rise to a $p(g|X,Y)$ which can be simulated in a classical model for the given scenario.

To this end, we focus on the a simple Bell scenario going beyond that of the CHSH inequality. In order to work with the smallest number of outcomes possible ($a,b\in[2]$), we must consider two parties with ternary settings  settings $x,y\in[3]$. Three settings are needed, as the two-setting scenario is fully characterised by the CHSH inequality, which is a correlation inequality and therefore implies advantages in a CCP whenever violated \cite{BZ04, BC01}. Alternatively, one could also consider the previously discussed Bell scenario with two settings and three outcomes. However, we focus on the former due to its conceptual simplicity and (as it turns out) computational advantages. In the three-setting scenario, the facet Bell inequalities are the (lifted) CHSH inequality and the $I_{3322}$ inequality \cite{Froissart, CollinsGisin}. The  $I_{3322}$ inequality reads
\begin{equation}\label{F}
I=-P_\text{A}(0)-2P_\text{B}(0)-P_\text{B}(1)+\sum_{x,y}T_{x,y}P(x,y)\leq 0,
\end{equation}
where $P(x,y)$ is the probability of outputting $a=b=0$, $P_\text{A}(x)=p(a=0|x)$,  $P_\text{B}(y)=p(b=0|y)$ and $T=\{[1,1,1],[1,1,-1],[1,-1,0]\}$. Notably, this inequality is not a correlation Bell inequality and is therefore not in the broad class of Bell inequalities whose violation necessarily implies advantages in CCPs \cite{BZ04}.

Motivated by the choice of scenario in previous discussions for translating Bell inequality violations to advantages in CCPs, we consider a communication scenario in which Alice has  inputs $x_0\in[2]$ and $x\in[3]$ and Bob has an input $y\in[3]$.  Alice sends  $m\in[2]$ to Bob who outputs $g\in[2]$. To further motivate that this scenario is a good choice for revealing the CCP advantages of probability distributions that violate the $I_{3322}$ inequality, we have shown in Appendix.~\ref{AppC} that a maximal violation of \eqref{F} (for qubits) implies better-than-classical communication complexity, and also that \textit{every} probability distribution violating \eqref{F} obtained from mixing the optimal one with a uniform probability distribution (in analogy with Eq.~\eqref{dist}) also implies such an advantage. Note that such a scenario is a natural extension of the ones studied in Ref.~\cite{BZ04}.

Nevertheless, we show that there exists a nonmaximally entangled state and local measurements that give rise to a probability distribution that violates $I_{3322}$ inequality, that however is  not advantageous in any CCP in the stated scenario. To this end, Alice and Bob can generate a Bell-like distribution of the form $p(a,b|x_0,x,y)$. Notice that only three of Alice's six inputs are required to create Bell nonlocal correlations that violate the $I_{3322}$ inequality. We can without loss of generality label these three inputs by $x$. Thus, with these labels, the dependence on $x_0$ in the Bell nonlocal distribution becomes trivial, i.e.~$p(a,b|x_0,x,y)=p(a,b|x,y)$. Now, we can choose our \textit{candidate probability distribution}  $p^{\text{cand}}(a,b\lvert x,y)$. This distribution has a quantum realisation. It also weakly violates Eq.~\eqref{F} ($I\approx 0.0129$), but importantly does not violate the CHSH inequality and hence cannot lead a better-than-classical score in a CCP based on the CHSH inequality. The candidate probability distribution was originally proposed in Ref.~\cite{CollinsGisin} and we detail it and its quantum realisation in
 Appendix.~\ref{AppD}.  We show that for every possible communication strategy within the scenario, there exists no CCP in which $p^{\text{cand}}$ enables an advantage over classical protocols. We first note that since we have fixed the probability distribution in the Bell scenario to $p^\text{cand}$, the set of distributions \eqref{quantum} that Alice and Bob can generate in the communication scenario forms a polytope. Therefore, it suffices to show that all deterministic communication strategies with access to $p^\text{cand}$ can be classically modelled. Since Alice maps the twelve possible values of $(a,x_0,x)$ to her binary message $m$, and Bob maps the twelve values $(m,b,y)$ to his binary output $g$, there exists a total of $2^{24}$ deterministic communication strategies. For each of these (indexed by $\mu$), we have evaluated the corresponding probability distribution $p_\mu(g\lvert x_0,x,y)=\sum_{a,b,m}p(a,b\lvert x,y)p_\mu(m\lvert a,x_0,x)p_\mu(g\lvert m,b,y)$. We have found that the relevant polytope of probability distributions in the communication scenario  has  $8192992$ vertices. To show that the probability distribution
 $p_\mu(g\lvert x_0,x,y)$ can be simulated by a classical model for all vertices, we consider the mixture of each vertex probability distribution with random outcomes; $p^{\text{Q},v}_\mu(g\lvert x_0,x,y)=vp_\mu(g\lvert x_0,x,y)+(1-v)/2$. Then, for each of the roughly eight million values of $\mu$, we decide the possibility of a classical model by running  
a linear program algorithm\footnote{Since each run of the linear program takes a few seconds to complete, it would require months to complete all $8192992$ cases on a standard computer. To contend with this problem, we have distributed the computation; roughly 40\% of it to the high-performance computing cluster 
\href{Baobabhttps://plone.unige.ch/distic/pub/hpc/baobab_en}{Baobab} 
at the University of Geneva, another 20\% to two workstation computers, and the remaining 40\% to five standard desktop computers. 
This allowed us to complete the full computation in less than three weeks.} 
 analogous to Eq.~\eqref{LP}. We find that for every choice of $\mu$, the value of $v$ is never smaller than one (up to machine precision). That is, every $p_{\mu}(g\lvert x_0,x,y)$ can be classically modelled. Thus, we conclude that  in the scenario in which Alice has $X\in[6]$  and Bob has $Y\in[3]$ and  $m$ and $g$ are bit valued, there exists no CCP that can be improved beyond classical constraints by the parties sharing the nonlocal probability distribution $p^\text{cand}(a,b\lvert x,y)$.

\section{Conclusions}  A substantial number of examples of quantum advantages in CCPs being powered by Bell inequality violations can be understood as different instances of a single map from Bell inequalities to CCPs. We found that a violation of the former implies an advantage in the latter for a simple class of communication strategies. As we explicitly showed, a complete analysis of classical communication complexity requires the revision of several previous claims in which violations of particular Bell inequalities where thought to imply advantages in CCPs. Going beyond that, we found that there exists nonlocal distributions for which the statistics of every possible communication strategy in any possible CCP can be simulated by classical models in an input/output scenario that naturally extends previous works. This suggests that not all forms of Bell nonlocality are useful for better-than-classical communication complexity. A definite proof of this statement would require an extension of our results to CCPs with any number of inputs and outputs. Our results motivate a characteriation of the (now seemingly nontrivial) relation between Bell nonlocality and entanglement-assisted communication complexity. Which nonlocal probability distributions are useful for outperforming classical limitations in CCPs and which are not?

\begin{acknowledgments}
We thank Fabian Bernards, Nicolas Brunner and Nicolas Gisin for discussions. We also thank Dmitry Tabakaev, Marc-Olivier Renou, Cai Yu, Sebastien Designolle and Davide Rusca for lending us their computers. Finally, we thank  Emmanuel Zambrini for much appreciated help with the Baobab computer system. AT acknowledges support from the Swiss National Science Foundation (Starting grant DIAQ, NCCR-QSIT). MZ acknowledges the ICTQT IRAP project of FNP, financed by structural funds of EU, and the COPERNICUS FNP/DFG grant-award. CB acknowledges the support of the Austrian Science Fund (FWF) through the SFB project "BeyondC" and the project I-2526-N27.

\end{acknowledgments}

\appendix

\section{The optimal classical score in the communication complexity problems of Refs.~\cite{BP03, TP17}}\label{AppA}
Ref.~\cite{BZ02} introduced CCPs based on the ternary-outcome CGLMP inequality. These were extended in Refs.~\cite{BP03, TP17} to CGLMP inequalities with any number ($d$) of outcomes. In a spirit similar to that of Ref.~\cite{BZ02}, these subsequent works restrict themselves to considering classical communication strategies of the additive type. Here, we show that also for $d>3$ this fails to capture the optimal classical score of the CCP. For simplicity, we focus on the case of four outcomes.

Refs.~\cite{BP03, TP17} present the following CCPs (up to minor modifications). Alice has eight possible inputs written in terms of a bit $x\in[2]$ and a quart $x_0\in[4]$. Bob has two possible inputs $y\in[2]$. Alice may communicate at most a four-valued message to Bob who aims to maximise the score
\begin{multline}\label{efficglmp}
\mathcal{S}_{\text{cglmp}}=\frac{1}{16}\sum_{x_0,x,y}P_{x,y}(g=x_0+f_1^1)-P_{x,y}(g=x_0+f_2^1)\\
+\frac{1}{3}\left(P_{x,y}(g=x_0+f_1^2)-P_{x,y}(g=x_0+f_2^2)\right),
\end{multline}
where 
\begin{align}\nonumber
& f_1^1=x_0-xy && f_2^1=x_0-xy+(-1)^{x+y}\\\nonumber
& f_1^2=x_0-xy-(-1)^{x+y} &&  f_2^2=x_0-xy+2(-1)^{x+y},
\end{align}
computed modulo four.  Alice (Bob) uses $x$ ($y$) to measure an entangled pair with possible outcomes $a\in[4]$ ($b\in[4]$). Then, using an additive communication strategy, i.e. $m=x_0+a\mod{4}$ and $g=m+b\mod{4}$, one finds that $\mathcal{S}_\text{cglmp}$ becomes identical to the Bell expression in the four-outcome CGLMP inequality \cite{BP03, TP17} which has an LHV bound (in this form and normalisation) of $1/2$. Therefore, under additive communication strategies Refs.~\cite{BP03, TP17} found $\mathcal{S}_\text{cglmp}\leq 1/2$.

However, the optimal classical score is not saturated with such a communication strategy. Since Alice maps eight inputs to four outputs, and similarly for Bob, there is a total of $4^{16}$ pairs of encoding and guessing functions. We have evaluated the score  for all such pairs and found that the optimal classical score is 
\begin{equation}
\mathcal{S}^\text{C}_{\text{cglmp}}=\frac{2}{3}.
\end{equation}
An encoding/decoding strategy that saturates this bound is
\begin{align}
m=[0,0,0,1,0,2,3,0]\\
g=[0,3,1,2,2,3,0,1],
\end{align}
where the tuple represents the response to the pair $(x,x_0)$ and $(y,m)$ respectively (ordered as $(0,0),(0,1),\ldots,(1,2),(1,3)$). Thus, in full analogy with the discussion in the main text focused on ternary-outcome CGLMP inequalities, corrections also apply to its generalisation to more than three outcomes.

\section{Numerical search for the optimal quantum score in the CCP based on the CGLMP inequality}\label{AppB}
We present numerical methods which we used
in support of the conjecture that an additive communication strategy and a quantum probability distribution that maximally violates the CGLMP inequality
give the optimal entanglement-assisted score in the CCP based on the ternary-outcome CGLMP inequality (\ref{score}).


The joint state of Bob's local system (after Alice's measurement) and the classical message when averaged over Alice's outcome can be written
\begin{equation}\label{B1}
\rho_{x_0x}=\sum_a \Tr_\text{A}\left[A_x^a\otimes \openone\rho\right]\otimes \ketbra{m}{m},
\end{equation}
where we encode the classical message in the computational basis state $\ketbra{m}{m}$. For all deterministic messages of the restricted class $m=m(a,x_0)$, we have evaluated the score and found that only those of the additive type lead to a better-than-classical score. However, for a general deterministic message $m=m(a,x_0,x)$ such a brute-force approach is too time-consuming. To address the general case, we can obtain upper bounds on the score by substituting $\ketbra{m}{m}$ in Eq.~\eqref{B1} with a quantum system $\sigma_{a,x_0,x}\in \mathbb{C}^3$. Notice that this only serves as a tool towards treating the relevant problem in which the message is classical. Moreover, this substitution is far more constraining than allowing for general quantum communication assisted by shared entanglement. The substitution of the classical message for a quantum one comes with the advantage that one can efficiently run alternating convex searches for lower bounding the quantity $\mathcal{S}_\text{cglmp}^\text{QC}=\max_{\rho,A,B,\sigma} \mathcal{S}_\text{cglmp}$. The searches are alternating in the sense that we first considers a semidefinite program optimising over the shared state, then a second one optimising over Alice's measurements, then a third one optimising over Bob's measurements and finally a fourth one optimising over the quantum message. This procedure of four semidefinite programs is iterated until the results appear to converge. For three respectively four dimensional entangled systems, we implemented the procedure by  alternating semidefinite programs each optimising over the state, Alice's measurements, Bob's measurements and the quantum message respectively. We have implemented this procedure for $10000$ randomly chosen starting points. Each of these $10000$ trials involves ten iterations of the described procedure (that is, 40 evaluations of a semidefinite program). In all $10000$ cases we find that the optimisation converges to the value $\mathcal{B}^\text{Q}_\text{cglmp}$, which is what is obtained by maximally violating the CGLMP inequality and then using an additive communication strategy.

\section{Communication complexity advantages via violation of the $I_{3322}$ inequality}\label{AppC}
Ref.~\cite{CollinsGisin} found that the maximal quantum violation of the $I_{3322}$ inequality with a shared entangled pair of qubits is $I^\text{Q}=1/4$ and is achieved with the singlet state $\ket{\psi^-}=(\ket{0,1}-\ket{1,0})/\sqrt{2}$ and measurements in the $xz$-plane of the Bloch sphere whose Bloch vectors (including the antipodal vectors) form a hexagon on both Alice's and Bob's side. Specifically, the Bloch vectors read
\begin{align}\nonumber
& \vec{a}_1=[0,0,1] && \vec{b}_1=-[\sqrt{3},0,1]/2\\\nonumber
& \vec{a}_2=[\sqrt{3},0,1]/2  && \vec{b}_2=-[0,0,1] \\
& \vec{a}_3=[\sqrt{3},0,-1]/2 && \vec{b}_3=[\sqrt{3},0,-1]/2.
\end{align}
Thus the resulting probability distribution is
\begin{equation}
p^\text{3322}(a,b\lvert x,y)=\frac{1}{4}\left[1-(-1)^{a+b}\vec{a}_x\cdot \vec{b}_y\right].
\end{equation}
We have considered the mixture of this probability distribution with a uniformly random probability distribution, i.e.
\begin{equation}\label{mix}
p^v(a,b\lvert x,y)=vp^\text{3322}(a,b\lvert x,y)+\frac{1-v}{4}.
\end{equation}
This probability distribution violates the $I_{3322}$ inequality only when $v>4/5$. We consider its usefulness in CCPs in a scenario (same as in the main text) in which Alice has six inputs, Bob has three inputs, and Alice's and Bob's outputs both are binary. We choose an additive communication strategy in which Alice sends $m=a+x_0\mod{2}$ and Bob outputs $g=m+b\mod{2}$. This leads to a specific probability distribution in the CCP (dependent on $v$). We then run a linear program of the type presented in the main text to determine the largest $v$ for which the entanglement-assisted probability distribution in the CCP has a classical model. We find that it returns $v=4/5$, thus showing that every probability distribution of the form \eqref{mix} that violates the $I_{3322}$ inequality implies advantages in a CCP.

\section{The candidate probability distribution}\label{AppD}
There exists probability distributions in the Bell scenario with three setting and two outcomes that violate the $I_{3322}$ inequality but not the CHSH inequality. Ref.~\cite{CollinsGisin} provided an example, which we in the main text used as the candidate probability distribution $p^{\text{cand}}(a,b\lvert x,y)$. It is obtained by Alice and Bob sharing the noisy state 
\begin{equation}
\rho=\frac{17}{20}\ketbra{\phi}{\phi}+\frac{3}{20}\ketbra{0,1}{0,1}
\end{equation}
where $\ket{\phi}=(2\ket{0,0}+\ket{1,1})\sqrt{5}$. This state cannot violate the CHSH inequality for any choice of measurements (which can be checked via the Horodecki criterion \cite{Horo}), but it can violate the $I_{3322}$ inequality as follows. Write Alice's (Bob's) Bloch vectors in the $xz$-plane as $\vec{a}_x=[\sin\theta_x,\cos\theta_x]$ ( $\vec{b}_y=[\sin\phi_y,\cos\phi_y]$) with
\begin{align}\nonumber
& \theta_1=\eta & \theta_2=-\eta && \theta_3=-\frac{\pi}{2}\\ \nonumber
& \phi_1=-\chi & \phi_2=\chi && \phi_3=\frac{\pi}{2}\\ \nonumber
\end{align}
with $\eta=\arccos\left(\sqrt{7/8}\right)$ and $\chi=\arccos\left(\sqrt{2/3}\right)$. This defines the candidate probability distribution
\begin{multline}
p^\text{cand}(a,b\lvert x,y)=\\
\frac{1}{4}\Tr\left[(\openone+(-1)^a\vec{a}_x\cdot \vec{\sigma})\otimes (\openone+(-1)^b\vec{b}_y\cdot \vec{\sigma})\rho\right]
\end{multline}
which achieves $I\approx 0.0129$.

\end{document}